\documentclass{mem}
\usepackage{natbib}\usepackage{txfonts}\usepackage{balance}
\usepackage{graphicx}
\usepackage[a4paper,breaklinks,dvips]{hyperref}
\idline{75}{282}

\begin{document}

\newcommand{\COBOLD}{{\tt CO$^5$BOLD}}
\newcommand{\LINFOR}{{\tt Linfor3D}}

\def\teff{$T\rm_{eff }$}
\def\kms{$\mathrm {km s}^{-1}$}

\title{Oxygen spectral line synthesis: 3D non-LTE with \COBOLD\ 
       hydrodynamical model atmospheres}

   \subtitle{}

\author{D.\,Prakapavi\v{c}ius\inst{1}, M.\,Steffen\inst{2,6},
A.\,Ku\v{c}inskas\inst{1, 3}, H.-G.\,Ludwig\inst{4}, B.\,Freytag\inst{5},
E.\,Caffau\inst{4,6}, R.\,Cayrel\inst{6}}

\institute{
Vilnius University Institute of Theoretical Physics and Astronomy,
A. Go\v{s}tauto 12,
Vilnius LT-01108, Lithuania
\and
Leibniz-Institut f\"{u}r Astrophysik Potsdam,
An der Sternwarte 16,
D-14482, Potsdam, Germany
\and
Vilnius University Astronomical Observatory,
M. K. \v{C}iurlionio 29,
Vilnius LT-03100, Lithuania
\and
ZAH Landessternwarte K\"{o}nigstuhl,
D-69117 Heidelberg, Germany
\and
Centre de Recherche Astrophysique de Lyon,
UMR 5574, CNRS, Universit\'{e} de Lyon,
\'{E}cole Normale Sup\'{e}rieure de Lyon,
46 All\'{e}e d'Italie,
F-69364 Lyon Cedex 07, France
\and
GEPI, Observatoire de Paris,
CNRS, UMR 8111, 61 Av. de l'Observatoire,
75014 Paris, France
}

\authorrunning{Prakapavi\v cius et al.}

\titlerunning{Oxygen spectral line synthesis in 3D-NLTE}

\abstract{In this work we present first results of our current project aimed 
  at combining the 3D hydrodynamical stellar atmosphere approach with non-LTE
  (NLTE) spectral line synthesis for a number of key chemical species. We carried 
  out a full 3D-NLTE spectrum synthesis of the oxygen IR 777\,nm triplet, using a
  modified and improved version of our NLTE3D package to calculate departure 
  coefficients for the atomic levels of oxygen in a \COBOLD\ 3D hydrodynamical solar
  model atmosphere. Spectral line synthesis was subsequently performed with
  the \LINFOR\ code. In agreement with previous studies, we find that the lines 
  of the oxygen triplet produce deeper cores under NLTE conditions, due to the diminished 
  line source function in the line forming region. This means that the solar oxygen 
  IR 777\,nm lines should be stronger in NLTE, leading to negative 3D~NLTE--LTE 
  abundance corrections. Qualitatively this result would support previous claims 
  for a relatively low solar oxygen abundance. Finally, we outline several further 
  steps that need to be taken in order to improve the physical realism and numerical 
  accuracy of our current 3D-NLTE calculations.}

\maketitle{}

\section{Introduction\label{Intro}}

Photospheric abundances of chemical elements derived from stellar spectra are
important means for testing and constraining models of the formation and
evolution of the Galaxy and its various stellar populations. The reliability of
the derived chemical abundances is limited, apart from the quality of the
observational data, by the realism of the ingredients used in the abundance
analysis: atomic data that describe the spectral lines themselves, stellar
model atmospheres, and the allowance of possible departures from local 
thermodynamic equilibrium (LTE).

During the recent years, reliable spectral line data has been measured and/or 
computed for a large number of astrophysically relevant spectral lines, and
departures from LTE are currently accounted for with an increasingly more 
realistic treatment of collisional processes. However, the stellar model 
atmospheres that are commonly used for the abundance analysis are based on 
a number of critical simplifications: they are typically constructed in 
one-dimensional (1D) geometry - plane-parallel or spherically-symmetric - 
and are subject to hydrostatic and radiative-convective equilibrium 
\citep[see, e.g.,][]{2004astro.ph..5087C, 2005ESASP.576..565B, 2008A&A...486..951G}. 
Even though the input microphysics (opacities, equation of state) of these 
models is sufficiently realistic, the convective energy transport, one of 
the main processes shaping the physical structure of the photosphere, is 
treated in the approximation of the mixing-length theory \citep{1958ZA.....46..108B} 
or its derivatives \citep{1991ApJ...370..295C}. 

In this respect, three-dimensional (3D) hydrodynamical stellar model atmospheres
are much more realistic, as they are able to account for convection based 
on first principles without the need of free parameters 
\citep[see][]{1982A&A...107....1N,1998ApJ...499..914S, 2000A&A...359..729A, 
2002AN....323..213F, 2004A&A...414.1121W}. Moreover, this type of model
atmospheres naturally allows for the emergence of a surface granulation
pattern, horizontal inhomogeneities and wave activity, all being unique
properties of 3D hydrodynamical model atmospheres. It has been found that 
the interplay of these physical processes significantly alters spectral 
line formation, which may lead to substantial differences in the abundances 
derived with 3D and 1D model atmospheres, respectively  \citep[][]
{2007A&A...469..687C, 2010nuco.confE.288D, 2010nuco.confE.290I, 2013A&A...549A..14K, DKS13}.
These findings conspicuously indicate that the 3D photospheric structure 
and dynamics has a significant impact on the spectral line formation, and 
should be properly taken into account in the chemical abundance analysis.

Besides the application of realistic stellar model atmospheres, 
an adequate treatment of non-LTE (NLTE) processes can be also very
important for deriving reliable chemical abundances. In the optically 
thin line-forming regions, the absorbing particles experience a radiation 
field that is of non-local origin since it forms deeper in the photosphere. 
Consequently, radiation field in the line-forming region may 
exhibit significant deviations from the local Planck function.
Radiation of non-local origin can distort the collisional ionization
balance given by the Saha equation, and drive the population numbers 
of the upper and lower atomic levels of the given transition away 
from the Boltzmann distribution that is valid in LTE \citep[for a 
thorough discussion of these effects see, e.g.][]{1992A&A...265..237B}.
Departures from LTE modify both the strength and shape of the 
spectral lines, and therefore can significantly alter spectroscopically
derived chemical abundances, in particular at low metallicities 
\citep[e.g.,][]{1999ApJ...521..753T}.

Given the significance of NLTE effects and the magnitude of the LTE 
3D-1D abundance corrections (especially in the metal-poor stars), it 
is obvious that the two factors should be simultaneously taken into 
account in order to derive reliable abundance estimates. However, 
the joint 3D-NLTE approach has only been applied in a few selected 
cases so far \citep{2004A&A...417..751A,
  2005ApJ...618..939S, 2007A&A...473L..37C, 2009A&A...508.1403P, 
  2010A&A...522A..26S, 2012MSAIS..22..152S}, therefore a systematic 
application of this admittedly very demanding methodology to, e.g., 
the investigation of stellar populations, is yet to come.

Driven by the need to combine modern 3D stellar model atmospheres and
non-LTE spectral line synthesis, we have recently started a project
to make this methodology available for a wider range of chemical elements
to be studied with 3D hydrodynamical \COBOLD\ stellar model atmospheres.
Oxygen is a particularly interesting element to investigate using the full 3D 
NLTE approach: it is the most abundant chemical element besides hydrogen
and helium, and its photospheric abundance is widely used to trace the
formation and chemical evolution of various Galactic populations.
It is well known that the \ion{O}{i} IR triplet ($\lambda = 777$~nm) experiences
significant departures from the LTE \citep[][and references therein]
{1993A&A...275..269K, 2009A&A...500.1221F}. Due to its high ionization
potential, oxygen does not experience significant overionization in 
late-type stars, so that NLTE effects are mainly limited to deviations 
from the thermal excitation equilibrium. These physical aspects limit 
the range of possible NLTE effects in the case of oxygen and therefore 
make it a good test case for the full 3D-NLTE approach with \COBOLD\ 
model atmospheres.

\section{Methodology\label{Method}}

  \subsection{Model atmospheres\label{Models}}

The 3D hydrodynamical solar model atmosphere used in this work was computed
with the \COBOLD\ code \citep{2012JCoPh.231..919F}. The \COBOLD\ code
solves time-dependent equations of compressible hydrodynamics and
radiation transfer on a 3D Cartesian grid. Computed in the "box-in-a-star"
setup \citep[for details see, e.g.,][]{2012JCoPh.231..919F}, our solar model
atmosphere was allowed to evolve hydrodynamically for several convective
turnover times \citep[see][Appendix A, for a definition of different time
scales]{2012A&A...547A.118L}. This particular \COBOLD\ simulation was also
used in the studies of \citet{2008A&A...488.1031C} and \citet{2012A&A...539A.121B},
to which we refer the reader for a more detailed description of the input
microphysics and physical properties of the model itself.

A representative sub-sample of twenty 3D model snapshots (i.e. model
structures calculated at different instants in time) was chosen out
of the relaxed part of the 3D model run in order to produce a
statistically representative and uncorrelated snapshot selection. More
specifically, snapshots in this sub-sample were chosen in such a way that
the average emerging radiation flux and its standard deviation would match
the corresponding values of the entire 3D model run. Similar requirements
were applied in the case of the horizontally averaged vertical mass flux at
characteristic optical depths. The final selection of the 3D model snapshots
obtained according to these criteria was subsequently used in the evaluation
of NLTE effects (Sect.~\ref{NLTE}) and, later, in the spectral synthesis 
calculations of the \ion{O}{i} IR triplet (Sect.~\ref{SpectSyn}).

The original model snapshots had 140 grid points in each horizontal
direction and 150 grid points on the vertical axis, corresponding
to a spatial coverage of $5.6\times5.6\times2.27$\,Mm. The vertical
grid of the model spanned the optical depth range of
$-6.7\mathrm{~<~log~\tau_{Ross}~<~}5.5$, which is sufficient to cover
the depths where the \ion{O}{i} IR triplet lines form. For the calculation
of departure coefficients and spectral line synthesis, a coarser 3D 
model atmosphere was constructed by choosing every third grid point 
horizontally, thereby reducing horizontal resolution of the model 
to $47\times47$ grid points. We have verified that differences in 
the spectral synthesis results obtained with the full and reduced 
model atmospheres, respectively, were negligible.

  \subsection{NLTE calculations\label{NLTE}}

Departures from LTE in the line formation computations are quantified
via three-dimensional sets of departure coefficients\footnote{The departure 
coefficient for atomic level $i=1\ldots i_{\rm max}$ is defined as
$b_i = N_{i,\,\mathrm{NLTE}}/N_{i,\,\mathrm{LTE}}$, where $N_i$ is the
level population for the respective case.} which in our study were 
computed with the NLTE3D code \citep[see][]{2007A&A...473L..37C,
2012MSAIS..22..152S}. We have recently made numerous improvements and 
generalizations to the code in order to adapt it for a wider variety 
of model atoms and astrophysical tasks. More specifically, we included 
IONDIS/OPALAM routines, originally part of the Kiel stellar atmosphere
package. These routines provide partition functions for a variety of 
chemical species and are also used for the computation of LTE population
numbers and continuous opacities at the wavelengths of the line transitions
in a given model atom. Since these routines are also used in the 
spectrum synthesis code \LINFOR\ (Sect.~\ref{SpectSyn}), this also 
contributes towards the self-consistency between the NLTE3D and 
\LINFOR\ packages.

During the first step, the NLTE3D code computes photo-ionization
and collisional rates, where the upward and downward collisional
rates are treated in detailed balance and, hence, only the
upward rates need to be computed explicitly. These rates are kept
constant throughout all further computations. After this step,
radiation transfer calculations are done for each of the considered
line transitions, $\ell=1\ldots \ell_{\rm max}$, to compute the mean
line intensity $\overline{J_\ell}(x,y,z)$ at all grid points, where $J$
refers to the angle-averaged intensity, and over-bar indicates averaging
over the line profile. Typically, we use 16 inclined plus the vertical 
direction for angular averaging, and 37 frequencies to resolve the
line profile. Then, the photo-excitation rates are computed using
$\overline{J_\ell}$, and all rates at each grid point of the 3D model
snapshot are passed to the statistical equilibrium routines that compute
NLTE population densities according to the prescriptions given
in \citet{1970stat.book.....M}. Given $\overline{J_\ell}$, a set of
linear equations needs to be solved for the unknown departure coefficients
$b_i, i=1\ldots i_{\rm max}$, which is done using standard linear algebra
routines, independently for each 3D model grid cell.

Once the new iteration of NLTE population numbers is completed,
the NLTE line opacities are updated and the next iteration is
started to compute line radiation transfer in order to obtain
updated photo excitation rates. This cycle is repeated via the
ordinary $\Lambda$-iteration scheme until the relative change
in the selected NLTE line equivalent widths (EWs) becomes
less than 10$^{-3}$ per iteration (see also Sect.~\ref{Converg}).

  \subsubsection{The oxygen model atom\label{Atom}}

  \begin{figure}
  \begin{center}
  \resizebox{\hsize}{!}{\includegraphics[clip=true,angle=0]{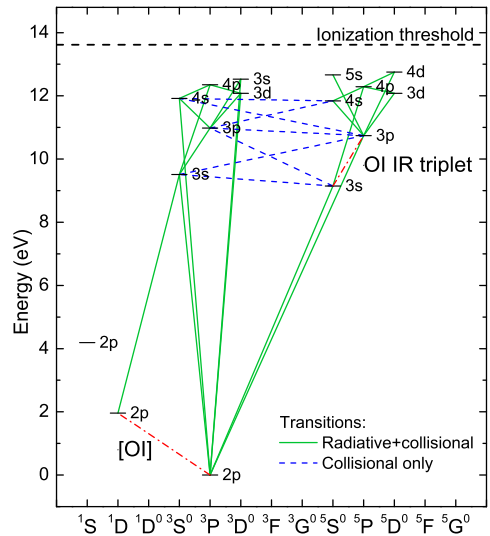}}
  \caption{Oxygen model atom used in this work. Radiatively allowed
transitions are marked by solid (green) lines, while radiatively forbidden
transitions that were treated only via electron-impact excitation are marked
by dashed (blue) lines. The forbidden [\ion{O}{i}] line and the \ion{O}{i}
IR triplet are marked by dash-dotted (red) lines. Collisional and radiative
bound-free transitions were taken into account for each level.}
  \label{Grotrian}
  \end{center}
  \end{figure}

The Grotrian diagram of the oxygen model atom used in our study is shown
in Fig.~\ref{Grotrian}. The model atom consists of $i_{\rm max}=16$ levels, 
connected by $16$ bound-free and $\ell_{\rm  max}=31$ bound-bound transitions.
Atomic data that describe the levels (energies, statistical weights) and 
radiative transitions (Einstein coefficients) were taken from the NIST
database. Photoionization cross-sections for the radiative bound-free 
transitions were adopted from the Topbase of the Opacity Project
\citep[][and further updates]{1993A&A...275L...5C}.

Excitation and ionization due to collisions with neutral hydrogen atoms was
treated via the \citet{1969ZPhy..225..470D} formula in the formulation of
\citet{1993PhST...47..186L}. In the present work, we set $S_{\rm H}$, a
parameter that scales the hydrogen collision rates, to 1.0.
Collisional ionization by electrons was treated using the classical
prescription of \citet{1962PPS....79.1105S}. Rate coefficients for the
excitation by inelastic electron collisions were taken from the work of 
\citet{2007A&A...462..781B}, allowing us to include the collisional
transitions between the triplet and quintet systems. Finally, we include
a resonant charge transfer reaction
$\mathrm{O^{0}~+~H^{+}~\rightleftharpoons~O^{+}~+~H^{0}}$ for the ground
level of oxygen, in the prescription by \citet{1985A&AS...60..425A}. The 
latter process is very efficient in late-type stars and therefore 
ensures that the ground level of \ion{O}{ii} is in LTE with respect 
to the ground level of \ion{O}{i}.

For the computation of the photo-ionizing radiation flux, opacity distribution
functions from \citet{2004astro.ph..5087C} were used, while the continuous
opacity was calculated with the IONDIS/OPALAM routines. Background opacities
stemming from the bound-bound transitions of elements other than oxygen
are not yet included in the current version of the NLTE3D code. Tests have
shown, however, that they are not important in the case of the oxygen model
atom used in this study.

In the statistical equilibrium calculations, we treated  the \ion{O}{i}
ground level (2p$^{3}$P) and the upper level of the \ion{O}{i} IR
triplet (3p$^{5}$P) as singlets, while, when performing spectrum synthesis
calculations, we have taken the fine-splitting into account, assuming
that the departure coefficients are identical for all sub-levels: due to
their close energetic values, they should be efficiently thermalized
and thus experience identical sensitivity to NLTE effects.

    \begin{figure*}
    \begin{center}
    \resizebox{11cm}{!}{\includegraphics{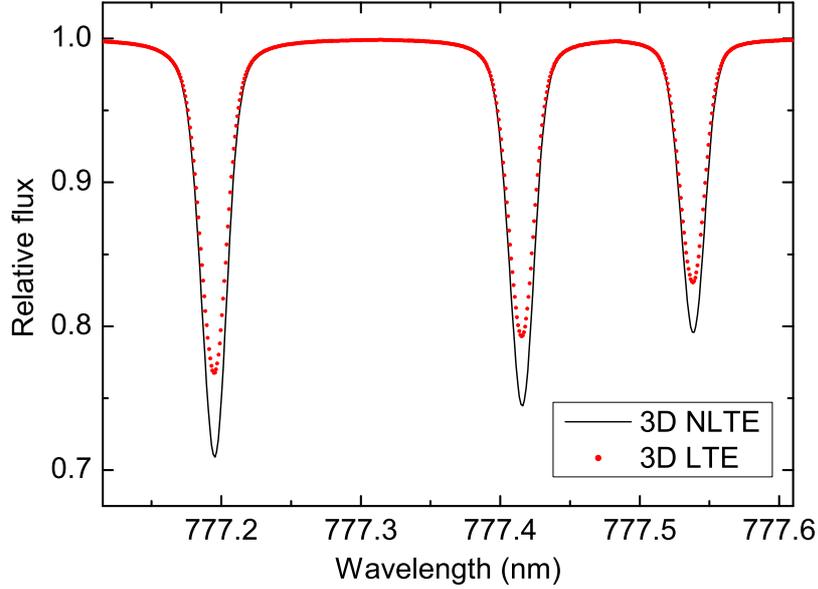}}
    \caption{Synthetic 3D NLTE (black line) and 3D LTE (red dots)
             \ion{O}{i} IR triplet lines in the solar spectrum
             calculated with identical oxygen abundance 
             (disk-integrated flux spectrum). Spectral line synthesis
             calculations were done using twenty 3D model snapshots 
             of a solar model atmosphere computed with the \COBOLD\ code.}
    \label{Spectrum}
    \end{center}
    \end{figure*}

  \subsection{Spectral synthesis calculations\label{SpectSyn}}

Spectral synthesis calculations of the \ion{O}{i} IR triplet lines were 
done with \LINFOR\footnote{http://www.aip.de/$\sim$mst/Linfor3D/linfor.pdf}.
Firstly, \LINFOR\ calculates the LTE line source function,
$S_{\rm LTE}(\nu)=B_\nu$ (Planck's black body function) and line opacity,
$\kappa_{\rm LTE}(\nu)$ for any given line transition. Whenever NLTE
spectrum synthesis is requested, both the line source function and the
line opacity are modified according to:
\begin{equation}
\frac{S_{\rm {NLTE}}(\nu)}{S_{\rm {LTE}}(\nu)} =
      \frac{\exp(\frac{h\nu}{kT})-1}
           {(b_{\rm low}/b_{\rm up}) \,\exp(\frac{h \nu}{kT})-1}\, ,
\label{eq:source}
\end{equation}
and
\begin{equation}
\frac{\kappa_{\rm {NLTE}}(\nu)}{\kappa_{\rm {LTE}}(\nu)} =
     b_{\rm low} \,\frac{\exp(\frac{h \nu}{kT}) - (b_{\rm up}/b_{\rm low}) }
           {\exp(\frac{h\nu}{kT})-1}\, ,
\label{eq:kappa}
\end{equation}
where $b_{\rm low}$ and $b_{\rm up}$ are the
departure coefficients of the lower and upper level of
the transition, respectively. Once the line source function
and line opacity are computed either in LTE or NLTE, the
radiation transfer is calculated along 16 inclined plus
the vertical direction, taking into account the differential
Doppler shifts along each line-of sight. This procedure is
repeated for every of the twenty 3D model snapshots. 
Finally, the resulting spectral lines synthesized
along each of the different directions are combined into a 
single spectral line profile by spatial, temporal, and angular 
averaging.

\section{Results\label{Results}}

  \subsection{Synthetic \ion{O}{i} IR triplet spectra\label{Synth}}

We have conducted NLTE and LTE spectrum synthesis calculations of the
\ion{O}{i} IR triplet lines using a 3D solar model atmosphere computed
with the \COBOLD\ code. Fig.~\ref{Spectrum} shows the obtained 
synthetic NLTE and LTE \ion{O}{i} IR triplet spectrum (identical 
oxygen abundances were used in the LTE and NLTE computations). It can
be immediately seen that both shape and strength of the \ion{O}{i}
IR triplet lines is sensitive to NLTE effects: while the line wings 
are very similar in both cases, deeper line core is present in 
NLTE. Hence, in accordance with the previous studies, NLTE effects 
lead to higher equivalent widths of the \ion{O}{i} IR triplet lines 
and to negative 3D~NLTE--LTE abundance corrections. Such behavior 
is qualitatively very similar to that reported by 
\citet{2004A&A...417..751A} or \citet{2009A&A...508.1403P}, which
was found to be one of the major factors leading to a lower 
solar oxygen abundance.

\begin{figure}
   \begin{center}
   \resizebox{\hsize}{!}{\includegraphics[bb=50 0 577 470,clip=true,angle=0]{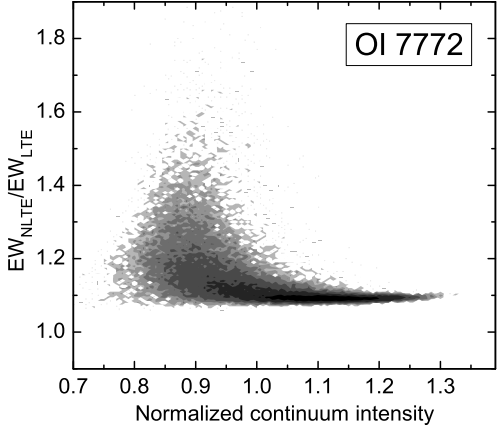}}
   \caption{Probability density plot showing the NLTE to LTE equivalent
            width ratio of the $777.19$~nm component of the \ion{O}{I}
            IR triplet, plotted against the normalized local continuum
            intensity for all twenty 3D model snapshots that were used
            in the calculations. Both continuum intensity and equivalent width 
            refer to the disk-center solar spectrum (vertical rays).}
   \label{EWIcont}
   \end{center}
   \end{figure}

The nature of the NLTE spectral line formation can be investigated by examining
the dependence of NLTE effects on the continuum intensity across the stellar
surface, where high and low continuum intensity represents granular and
intergranular regions, respectively. Fig.~\ref{EWIcont} shows the
distribution of the NLTE to LTE ratios of local EWs computed from the vertical
ray (disk-center intensity) and plotted versus the local relative continuum
intensity. While the plot reveals significant differences with respect to
the results obtained by \citet[][Fig. 4]{2004A&A...417..751A}, who found
diminished NLTE EWs in some (tough not all) intergranular lanes, our results
show a quite good qualitative agreement with the computations of 
\citet[][Fig. 8]{1993A&A...275..269K} (note, however, that the adopted 
$S\mathrm{_{H}}$ values were different in all three investigations). Our
results indicate that, in NLTE, line strengths are enhanced at every position 
on the disk and NLTE effects are more pronounced in the intergranular regions 
(low continuum intensity). \citet{1993A&A...275..269K}
suggested that this happens because in the intergranular regions the triplet
lines form higher in the atmosphere where the departures from LTE are more severe.

\begin{figure}
  \begin{center}
  \resizebox{\hsize}{!}{\includegraphics[bb=28 0 577 550,clip=true,angle=0]{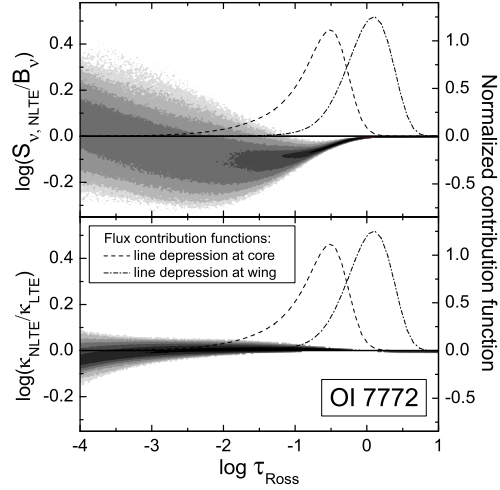}}
   \caption{NLTE-to-LTE ratio of the line source function (top panel)
            and line opacity (bottom panel) plotted versus Rosseland
            optical depth (gray scales represent the probability density).
            Both panels contain the same normalized flux contribution functions
            for the line depression at the core ($777.1954$~nm, dashed) and
            wing ($777.15$~nm, dot-dashed) of the \ion{O}{i} 7772 line.}
   \label{S_kappa}
   \end{center}
   \end{figure}

The origin of the NLTE effects can
be examined by comparing NLTE and LTE line source functions and line
opacities \citep[see, e.g., the analysis of][]{1993A&A...275..269K}.
Fig.~\ref{S_kappa} shows the ratio of NLTE and LTE line source functions
(top panel) and line opacities (bottom panel) for the 777.19 nm component
of the \ion{O}{i} IR triplet, as given by Equations \,(\ref{eq:source})
and (\ref{eq:kappa}). As was stated previously, the 3p$^{5}$P
sub-levels share identical departure coefficients, so the factor
$b_{\rm low}/b_{\rm up}$ that enters the equations and hence the ratios
$S_{\rm {NLTE}}/S_{\rm {LTE}}$ and $\kappa_{\rm {NLTE}}/ \kappa_{\rm {LTE}}$
are identical for each component of the triplet.

The plot shows that the source function decreases below the local Planck
function at $\mathrm{\log\tau_{\rm Ross}\lesssim0.0}$. Fig.~\ref{S_kappa}
also contains two spatially and temporally averaged flux contribution
functions for the line depression \citep[see][]{1986A&A...163..135M}
that were computed at the line center (777.1954 nm) and wing (777.15 nm)
of the line profile. It can be seen that the line wings form in a region
where $S_{\rm NLTE}$ is identical or very similar to $B_\nu$, while
the line core forms where significant departures from 
LTE may occur.

On the other hand, the line opacity of the \ion{O}{i} IR triplet stays
close to LTE in the entire line forming region. The NLTE/LTE ratio of
line opacities depends mostly on $b_{\rm low}$, which is close to unity
in the entire optical depth range where the \ion{O}{i} IR triplet lines
form, while the sensitivity to $b_{\rm low}/b_{\rm up}$ ratio is 
rather small. Hence, the influence of NLTE effects on the line opacity 
is weak, and therefore it can be inferred that the strengthening of 
the \ion{O}{i} IR triplet happens solely due to the diminished line 
source function. According to \citet[][and references therein]
{2009A&A...500.1221F}, this is caused by the photon losses in the line.

  \subsection{Numerical issues\label{Converg}}

\begin{figure}
  \begin{center}
  \resizebox{\hsize}{!}{\includegraphics[clip=true,angle=0]{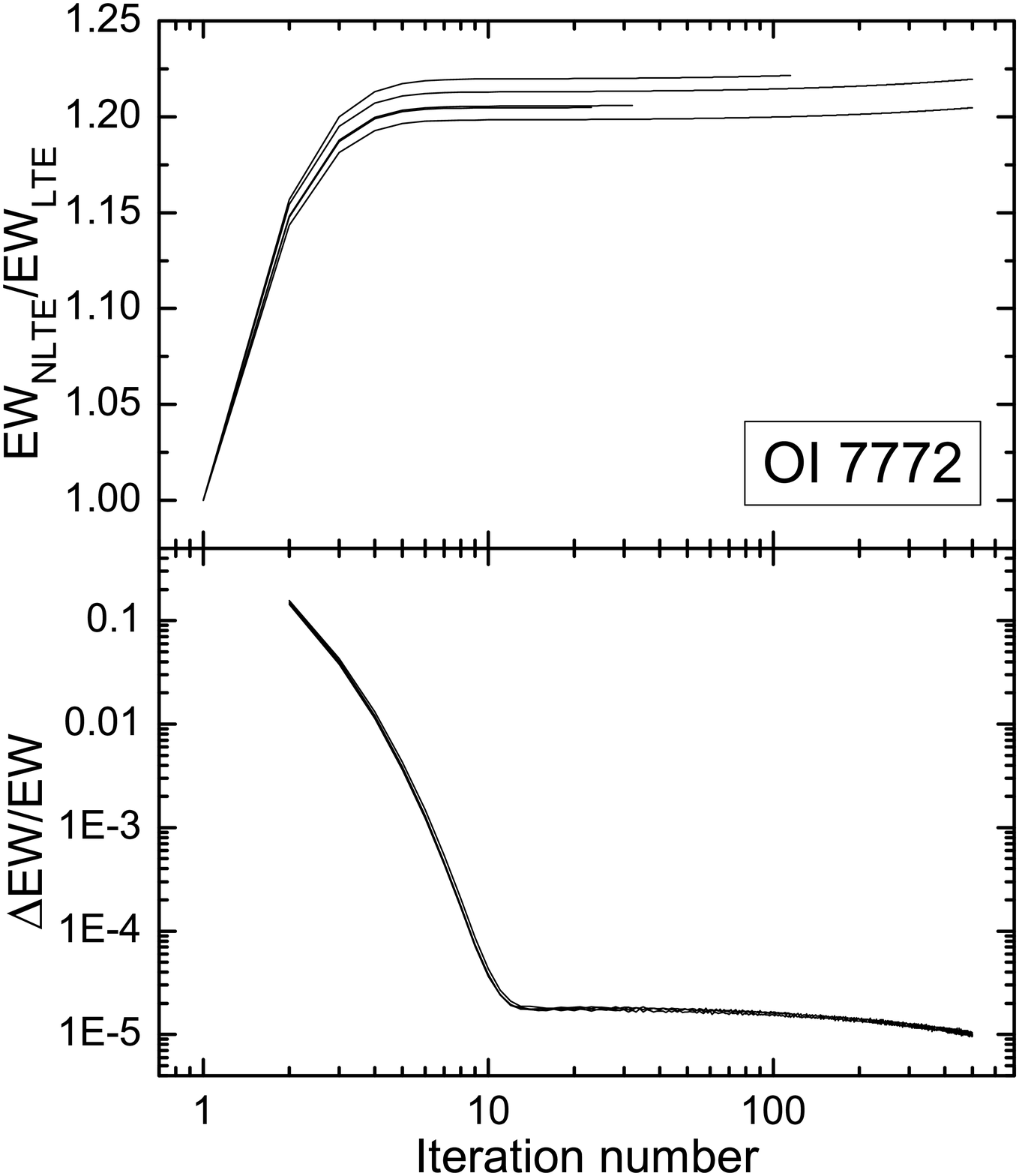}}
  \caption{Top panel: evolution of the disk-integrated equivalent width (EW)
  of the \ion{O}{i} 777.159 nm line during the iteration process. Bottom
  panel: change per iteration in the EW of a given \ion{O}{i} triplet component.
  The figure shows the iteration progress in five individual 3D model snapshots.
  Note that in the bottom panel data from the individual snapshots merge into
  one single curve.}
  \label{Convergence}
  \end{center}
  \end{figure}

Fig.~\ref{Convergence} shows the evolution of the equivalent width
of the $777.19$~ nm line (top panel) and the relative change
of EW (bottom panel) as a function of the iteration number. As
stated previously, the $\Lambda$-iterations are continued until the 
relative change in all NLTE equivalent widths of the selected radiative 
transitions (green lines in Fig.~\ref{Grotrian}) becomes less than 
10$^{-3}$ per iteration. Even though the triplet line shown in 
Fig.~\ref{Convergence} satisfies the convergence criterion already
after a small number of iterations, certain other lines (mainly the
resonant UV lines, not shown in the figure) may show much slower 
convergence and may thus require significantly more iterations to 
obtain the final NLTE population densities.

The top panel of Fig.~\ref{Convergence} shows that there is a wide
range in the number of iterations required to arrive at the 
convergence limit: depending on the selected 3D model snapshot,
it may be as low as $\sim30$ or as large as $\sim500$ (in the most
extreme cases the EWs of \ion{O}{i} IR triplet lines continue to
grow even after $\sim$300 iterations). One should note, however,
that the poor convergence properties of the ordinary $\Lambda$-iteration
scheme pose a well-known problem in radiation transfer calculations.
To quote \citet{2003ASPC..288...17H}: ``it exhibits a pathological
behavior in that the solution stabilizes (i.e. relative changes of
the source function become extremely small) long before the correct
solution is reached''. All this indicates that a more efficient
iteration scheme needs to be implemented into the NLTE3D code in
order to make the computations more efficient and accurate.

\section{Conclusions\label{Concl}}

We provide a short summary of our first results regarding the 3D NLTE 
spectrum synthesis of the \ion{O}{i} IR triplet lines, carried out 
using a 3D hydrodynamical solar model atmosphere computed with 
the \COBOLD\ code. Atomic level population numbers were calculated 
using a significantly expanded and improved version of the NLTE3D code. 
Our results show that, in agreement with previous studies, the \ion{O}{i} 
IR triplet lines are sensitive to NLTE effects, which lead to deeper line 
cores than those expected in LTE. This happens mainly because in the range
of optical depths where the cores of the \ion{O}{i} IR triplet lines form
the line source function becomes smaller than the local Planck function.
A significant deepening of the core, along with a net strengthening of
the equivalent width of the line, leads to negative 3D~NLTE--LTE
abundance corrections.

The current version of the NLTE3D code utilizes an ordinary $\Lambda$-iteration
scheme for the solution of non-linear statistical equilibrium equations.
Such approach may be adequate in cases where all line transitions are weak
and thus the radiative rates in the lines are only weakly coupled to the
level populations (e.g.\ lithium). This, however, is obviously not true in 
the case of oxygen. For a proper treatment of this more general situation, 
the implementation of a faster and more reliable iteration scheme via some 
form of accelerated $\Lambda$-iteration 
\citep[e.g.][]{1986JQSRT..35..431O, 1991A&A...245..171R}
may be needed. This will be amongst the main priorities for the
further development of the NLTE3D code, in order to ensure a higher 
reliability of the numerical results and a better versatility of the code 
in further applications involving more complicated model atoms.

\begin{acknowledgements}
{This work was supported by grant from the Research Council of Lithuania MIP-101/2011. 
 DP acknowledges financial support from Deutscher Akademischer Austausch
 Dienst (DAAD) that allowed exchange visits between Vilnius and Potsdam. 
 MS acknowledges funding from the Research Council of Lithuania for a research 
 visit to Vilnius. HGL, and EC acknowledge financial support by the 
 Sonderforschungsbereich SFB\,881 ``The Milky Way System'' (subproject A4) of 
 the German Research Foundation (DFG).}
\end{acknowledgements}

\bibliographystyle{aa}

\end{document}